\documentclass[a4paper,10pt]{article}
\usepackage{geometry}
\usepackage{amsmath,amssymb,amsfonts}
\usepackage{graphicx}
\usepackage{caption}
\usepackage{xcolor}
\usepackage{hyperref}

\usepackage{rotating}
\usepackage{graphicx}
\usepackage{array}
\usepackage{adjustbox}
\usepackage{lscape}

\usepackage{tabularx}
\usepackage{makecell}

\title{Introducing JIRIAF: A Virtual Kubelet Integration for Optimizing HPC Resource Provisioning}
\author{Vardan Gyurjyan, Graham Heyes, Christopher Larrieu, David Lawrence, Jeng-Yuan Tsai}
\date{}

\begin{document}

\maketitle

\begin{abstract}
The JIRIAF (JLab Integrated Research Infrastructure Across Facilities) framework is designed to streamline resource management and optimize high-performance computing (HPC) workloads across heterogeneous environments. Central to JIRIAF is the JIRIAF Resource Manager (JRM), which effectively leverages Kubernetes and Virtual Kubelet to manage resources dynamically, even in environments with restricted user privileges. By operating in userspace, JRM facilitates the execution of user applications as containers across diverse computing sites, ensuring unified control and monitoring. The framework's effectiveness is demonstrated through a case study involving the deployment of data-stream processing pipelines on the Perlmutter system at NERSC, showcasing its capability to manage large-scale HPC applications efficiently. Additionally, we discuss the integration of a digital twin model for a simulated queue system related to a streaming system, using a Dynamic Bayesian Network (DBN) to enhance real-time monitoring and control, providing valuable insights into system performance and optimization strategies.

\end{abstract}

\tableofcontents
\newpage

\section{Introduction}
High-performance computing (HPC) environments are increasingly heterogeneous, comprising various types of computing resources such as CPUs, GPUs, and specialized accelerators spread across multiple facilities. Efficiently managing and utilizing these diverse resources is crucial for maximizing computational throughput and minimizing operational costs. Traditional resource management frameworks often struggle to adapt to the complexities and dynamic nature of modern HPC environments, particularly when dealing with user privilege restrictions and varying resource availability.

The JLab Integrated Research Infrastructure Across Facilities (JIRIAF) framework addresses these challenges by providing a scalable and flexible resource management solution designed to seamlessly integrate and manage heterogeneous HPC resources. Central to JIRIAF is the JIRIAF Resource Manager (JRM), which plays a pivotal role in this framework. The JRM leverages the Kubernetes framework with Virtual Kubelet to enable resource management in environments where traditional kubelet installation is impractical due to user privilege restrictions. By operating in userspace, the Virtual Kubelet implementation allows JIRIAF to execute user applications as containers across different computing sites, providing unified control and monitoring capabilities.

This paper presents the design and implementation of the JIRIAF framework, with a particular emphasis on the JIRIAF Resource Manager (JRM). We demonstrate the framework's effectiveness through a proof-of-concept deployment on the Perlmutter system at the National Energy Research Scientific Computing Center (NERSC). The deployment involves data-stream processing pipelines for the CLAS12 experiment, showcasing JIRIAF's ability to manage large-scale HPC applications efficiently. Additionally, we discuss the integration of a digital twin model for a simulated queue system related to a streaming system. The digital twin, implemented using a Dynamic Bayesian Network (DBN), enhances real-time monitoring and control, offering insights into system performance and optimization strategies.

\section{Motivation}
The primary motivation behind JIRIAF is to streamline the management of large-scale distributed infrastructures. Key challenges include:
\begin{itemize}
    \item Efficiently migrating and scaling workloads across multiple computing sites.
    \item Intelligently utilizing opportunistic resources to enhance overall efficiency.
    \item Maintaining system integrity while operating in user space.
\end{itemize}

\section{Architecture}
\begin{figure}[htbp]
\centerline{\includegraphics[width=\linewidth]{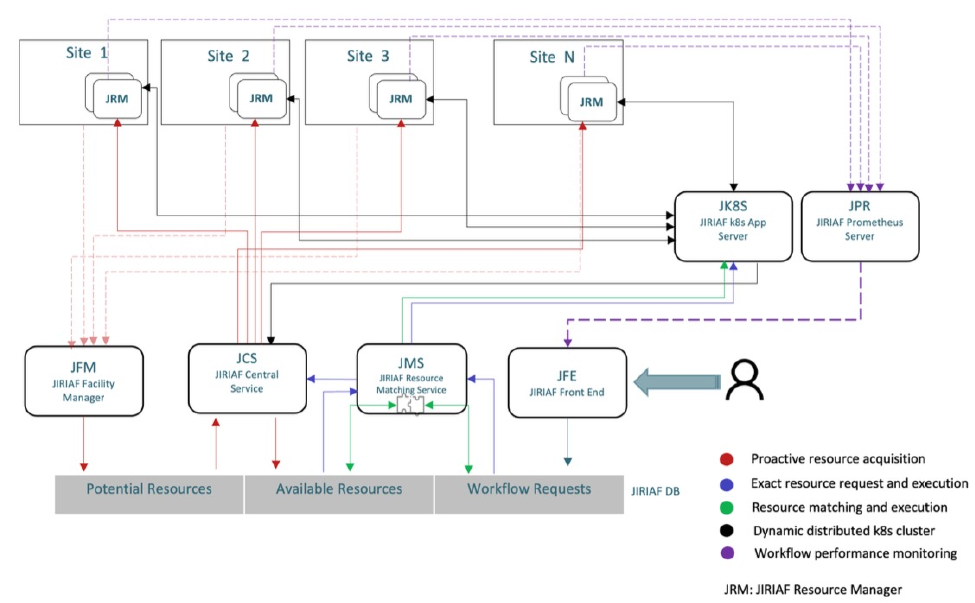}}
\caption{JIRIAF System Architecture and Workflow: This figure visually represents the sophisticated architecture of JIRIAF, emphasizing the roles and interconnectivity of its key components — the JFM, JCS, JRM, JMS, and JFE. By illustrating the flow of data and control across these components, the diagram elucidates the dynamic and efficient resource management system designed to optimize high-performance computing across heterogeneous environments.}
\label{arch}
\end{figure}

The JIRIAF architecture is meticulously designed to enable seamless integration and efficient resource management across diverse computing facilities. At the heart of this system is the JFM (JIRIAF Facility Manager), responsible for maintaining a dynamic resource pool by periodically scraping data from each computing facility to ensure an up-to-date inventory of available resources. The JCS (JIRIAF Central Service) functions as the central command, initiating pilot jobs through the JRM (JIRIAF Resource Manager). The JRM, which can operate in userspace to accommodate heterogeneous HPC setups, leases resources reported by the JFM, awaiting utilization. The JMS (JIRIAF Matching Service Algorithm) then steps in to update the available resource database, aligning resources with user requests. The JFE (JIRIAF Front End) finalizes the process by managing user requests and populating the user workflow request table. This comprehensive architecture is depicted in Figure~\ref{arch}, illustrating JIRIAF's commitment to providing a seamless and efficient computing environment.

\newpage

\section{Core Component - JIRIAF Resource Manager}
The JIRIAF Resource Manager (JRM) serves as an integral component of the framework, effectively leveraging the Kubernetes framework with Virtual Kubelet \cite{virtual-kubelet}. A fundamental Kubernetes cluster comprises a master node/control plane that oversees cluster management and worker nodes with kubelets connected to the containerd socket for container execution. Given that installing a regular kubelet necessitates root credentials, which are typically unavailable to ordinary users at compute sites, we employ the Virtual Kubelet to circumvent this limitation while still utilizing the Kubernetes framework. The Virtual-Kubelet-Cmd (VK) is a virtual kubelet implemented using BASH commands and operates in userspace. It translates a container into a BASH script composed of several processes. This approach serving as JRM allows JIRIAF to execute user applications as containers across various computing sites by simply running BASH commands in userspace, all the while ensuring unified control and monitoring through Kubernetes.

\subsection{Starting Virtual-Kubelet-Cmd}
Virtual-Kubelet-Cmd (VK) is a critical component that must be properly initialized to function within a Kubernetes environment. This section delineates the procedures for initializing VK using either the VK binary or a Docker image. The source code is accessible in the GitHub repository \cite{virtual-kubelet-cmd}.

\subsubsection{Using VK Binary to Start Virtual-Kubelet-Cmd}

The following script initializes Virtual-Kubelet-Cmd (VK) with the environment variables listed in Table \ref{table:environment_variables}:

\begin{verbatim}
#!/bin/bash
export MAIN="/workspaces/virtual-kubelet-cmd"
export VK_PATH="$MAIN/test-run/apiserver"
export VK_BIN="$MAIN/bin"
export APISERVER_CERT_LOCATION="$VK_PATH/client.crt"
export APISERVER_KEY_LOCATION="$VK_PATH/client.key"
export KUBECONFIG="$HOME/.kube/config"
export NODENAME="vk"
export VKUBELET_POD_IP="172.17.0.1"
export KUBELET_PORT="10255"
export JIRIAF_WALLTIME="60"
export JIRIAF_NODETYPE="cpu"
export JIRIAF_SITE="Local"

"$VK_BIN/virtual-kubelet" --nodename $NODENAME --provider mock --klog.v 3 > ./$NODENAME.log 2>&1
\end{verbatim}

\begin{table}[h!]
\centering
\caption{Environment Variables used to start VK}
\begin{tabular}{|l|p{10cm}|}
\hline
\textbf{Environment Variable} & \textbf{Description} \\
\hline
\texttt{MAIN} & Main workspace directory \\
\hline
\texttt{VK\_PATH} & Path to the directory containing the apiserver files \\
\hline
\texttt{VK\_BIN} & Path to the binary files \\
\hline
\texttt{APISERVER\_CERT\_LOCATION} & Location of the apiserver certificate \\
\hline
\texttt{APISERVER\_KEY\_LOCATION} & Location of the apiserver key \\
\hline
\texttt{KUBECONFIG} & Kubernetes configuration file location (default: \texttt{\$HOME/.kube/config}) \\
\hline
\texttt{NODENAME} & Name of the node in the Kubernetes cluster \\
\hline
\texttt{VKUBELET\_POD\_IP} & IP address of Pod IP \\
\hline
\texttt{KUBELET\_PORT} & Kubelet HTTP server port (default: 10250) \\
\hline
\texttt{JIRIAF\_WALLTIME} & Node runtime limit in seconds (0 for no limit) \\
\hline
\texttt{JIRIAF\_NODETYPE} & Node type for labeling purposes \\
\hline
\texttt{JIRIAF\_SITE} & Site for labeling purposes \\
\hline
\end{tabular}
\label{table:environment_variables}
\end{table}

\subsubsection{Using Docker Image for Virtual-Kubelet-Cmd}

The Docker image for \texttt{virtual-kubelet-cmd} is available on Docker Hub \cite{docker-vk-cmd}. This image includes the \texttt{virtual-kubelet-cmd} binary and necessary files, organized into a \texttt{vk-cmd} directory within the container. The containerization of VK is inspired by the KinD project \cite{KinD}, and the source code can be found on GitHub \cite{vk-cmd-build-img}.

The following script demonstrates how to run VK using the Docker image:

\begin{verbatim}
#!/bin/bash
export NODENAME="vk"
export KUBECONFIG="$HOME/.kube/config"
export VKUBELET_POD_IP="123.123.123.123"
export KUBELET_PORT="10260"

export JIRIAF_WALLTIME="0"
export JIRIAF_NODETYPE="cpu"
export JIRIAF_SITE="mylin"

export VK_CMD_IMAGE="jlabtsai/vk-cmd:main"
docker pull $VK_CMD_IMAGE

container_id=$(docker run -itd --rm --name vk-cmd $VK_CMD_IMAGE)
docker cp ${container_id}:/vk-cmd `pwd` && docker stop ${container_id}
cd `pwd`/vk-cmd

./start.sh $KUBECONFIG $NODENAME $VKUBELET_POD_IP $KUBELET_PORT $JIRIAF_WALLTIME \
    $JIRIAF_NODETYPE $JIRIAF_SITE
\end{verbatim}

When running VK on a remote machine, it is necessary to establish SSH tunnels to the control plane. Two tunnels are required: one for the API server and one for the metrics server. The following commands establish these tunnels on the worker node:

\begin{table}[h!]
\centering
\caption{SSH Tunnel for VK to API server}
\begin{tabular}{|l|l|}
\hline
\textbf{Connection} & API server listens to VK \\
\hline
\textbf{Environment Variables} & \texttt{APISERVER\_PORT}, \texttt{CONTROL\_PLANE\_IP} \\
\hline
\textbf{SSH Command} & \texttt{ssh -NfL \$APISERVER\_PORT:localhost:\$APISERVER\_PORT \$CONTROL\_PLANE\_IP} \\
\hline
\end{tabular}
\label{table:ssh_tunnel_vk_to_api}
\end{table}

\begin{table}[h!]
\centering
\caption{SSH Tunnel for Metrics server to VK}
\begin{tabular}{|l|l|}
\hline
\textbf{Connection} & VK listens to metrics server \\
\hline
\textbf{Environment Variables} & \texttt{KUBELET\_PORT}, \texttt{CONTROL\_PLANE\_IP} \\
\hline
\textbf{SSH Command} & \texttt{ssh -NfR *:\$KUBELET\_PORT:localhost:\$KUBELET\_PORT \$CONTROL\_PLANE\_IP} \\
\hline
\end{tabular}
\label{table:ssh_tunnel_metrics_to_vk}
\end{table}

\textbf{Note:} The \texttt{*} symbol in the SSH command for the metrics server to VK is used to bind the port to all network interfaces on the control plane.

\subsection{Deploying and Managing Pods on VK Nodes}
\label{deploying-pods}

Pods and their containers can be deployed on VK nodes. Table \ref{table:vk_vs_kubelet} provides a comparative analysis of VK nodes and standard kubelets:

\begin{table}[h!]
\centering
\caption{Comparison of VK Nodes and Regular Kubelets}
\begin{tabular}{|l|l|l|}
\hline
\textbf{Property} & \textbf{Virtual-Kubelet-CMD} & \textbf{Regular Kubelet} \\
\hline
Container Execution & Executes as Linux processes & Runs as a Docker container \\
\hline
Image Definition & Defined as a BASH script & Defined as a Docker container image \\
\hline
\end{tabular}
\label{table:vk_vs_kubelet}
\end{table}

\subsubsection{Managing Container Processes Using Process Group ID}

The \texttt{pgid} file records the process group ID (PGID) of the processes initiated by the script. Each container has a unique \texttt{pgid} file, located at \texttt{\$HOME/\$podName/containers/\$containerName/pgid}. This file is essential for controlling and managing the process group, ensuring that all processes within the group are appropriately handled.

\subsubsection{Procedure to Deploy a Pod Executing a BASH Script}

To deploy a pod that runs a script, follow these steps:

\begin{itemize}
    \item Store the script \texttt{stress.sh} in a \texttt{ConfigMap}.
    \item Mount the script into the container using \texttt{volumeMounts}. The \texttt{stress.sh} script will be located at \texttt{\$HOME/\$podName/containers/direct-stress/stress/} when the pod starts running.
    \item Specify the script execution using the \texttt{command} and \texttt{args} fields.
\end{itemize}

An example of a pod running a shell script is provided below:

\begin{verbatim}
kind: ConfigMap
apiVersion: v1
metadata:
  name: direct-stress
data:
  stress.sh: |
    #!/bin/bash
    stress --timeout $1 --cpu $2
---
apiVersion: v1
kind: Pod
metadata:
  name: direct-stress
spec:
  containers:
    - name: direct-stress
      image: direct-stress
      command: ["bash", "/stress/stress.sh"]
      args: ["300", "2"]
      volumeMounts:
        - name: direct-stress
          mountPath: /stress
  volumes:
    - name: direct-stress
      configMap:
        name: direct-stress
  nodeSelector:
    kubernetes.io/role: agent
  tolerations:
    - key: "virtual-kubelet.io/provider"
      value: "mock"
      effect: "NoSchedule"
\end{verbatim}

Important parameters in the YAML file are outlined in Table \ref{table:script_storage_execution}.

\begin{table}[h!]
\centering
\caption{Features for Script Management and Execution in Pods}
\begin{tabular}{|l|p{10cm}|}
\hline
\textbf{Feature} & \textbf{Description} \\
\hline
\texttt{ConfigMap} & Volume type for storing the script \\
\hline
\texttt{volumes} & Manages the use of \texttt{ConfigMap} \\
\hline
\texttt{volumeMounts} & Relocates the script to the specified path as \texttt{\$HOME/\$podName/containers/\$containerName/\$mountPath/} \\
\hline
\texttt{command} and \texttt{args} & Specifies the script to be executed and its arguments \\
\hline
\texttt{env} & Passes environment variables to the script \\
\hline
\texttt{image} & Name of the container image, same as the volume name for the script \\
\hline
\end{tabular}
\label{table:script_storage_execution}
\end{table}

\subsubsection{Setting Affinity for Pods on VK Nodes}

Pod affinity for VK nodes is determined by three labels: \texttt{jiriaf.nodetype}, \texttt{jiriaf.site}, and \texttt{jiriaf.alivetime}. These labels correspond to the variables \texttt{JIRIAF\_NODETYPE}, \texttt{JIRIAF\_SITE}, and \texttt{JIRIAF\_WALLTIME} in the \texttt{start.sh} script.

\begin{itemize}
    \item If \texttt{JIRIAF\_WALLTIME} is set to \texttt{0}, the \texttt{jiriaf.alivetime} label is not defined, and thus, affinity is not applied.
    \item VK status changes from \texttt{Ready} to \texttt{NotReady} once \texttt{jiriaf.alivetime} reaches zero. However, the VK process is not terminated.
    \item To add additional labels to VK nodes, modify the \texttt{ConfigureNode} function in \texttt{internal/provider/mock/mock.go} in the source code of GitHub \cite{virtual-kubelet-cmd}.
\end{itemize}

An example of setting affinity is provided below:

\begin{verbatim}
affinity:
  nodeAffinity:
    requiredDuringSchedulingIgnoredDuringExecution:
      nodeSelectorTerms:
      - matchExpressions:
        - key: jiriaf.nodetype
          operator: In
          values:
          - "cpu"
        - key: jiriaf.site
          operator: In
          values:
          - "nersc"
        - key: jiriaf.alivetime
          operator: Gt
          values:
          - "10"
\end{verbatim}

\subsection{Lifecycle of Containers and Pods}
\label{lifecycle}

The lifecycle of containers and pods defined by VK involves various states and transitions, crucial for monitoring and managing workload execution effectively. This section describes the container states and the methods used to create and monitor them.

\subsubsection{Description of Container States}

Each field in the container state tables provides specific information about the container's status and transitions, as described in Table \ref{table:field_descriptions}. Container states represent the stages a container goes through during its lifecycle. These states are essential for understanding the behavior and status of containers as they are created, run, and terminated. Notice that these methods can be found in \texttt{internal/provider/mock/mock.go} in the source code of GitHub \cite{virtual-kubelet-cmd}.

\paragraph{Container State Transitions in \texttt{CreatePod} Method}

The \texttt{CreatePod} method transitions the container through several states. Table \ref{table:uid_index_createpod} lists the UID Index for these states.

\begin{table}[h!]
\centering
\caption{UID Index for \texttt{CreatePod} method}
\label{table:uid_index_createpod}
\begin{tabular}{|l|l|}
\hline
\textbf{UID} & \textbf{UID Index} \\
\hline
create-cont-readDefaultVolDirError & 0 \\
\hline
create-cont-copyFileError & 1 \\
\hline
create-cont-cmdStartError & 2 \\
\hline
create-cont-getPgidError & 3 \\
\hline
create-cont-createStdoutFileError & 4 \\
\hline
create-cont-createStderrFileError & 5 \\
\hline
create-cont-cmdWaitError & 6 \\
\hline
create-cont-writePgidError & 7 \\
\hline
create-cont-containerStarted & 8 \\
\hline
\end{tabular}
\end{table}

\paragraph{Container State Transitions in \texttt{GetPods} Method}

The \texttt{GetPods} method periodically monitors the container's status. Table \ref{table:uid_index_getpods} lists the UID Index for the \texttt{GetPods} method.

\begin{table}[h!]
\centering
\caption{UID Index for \texttt{GetPods} method}
\label{table:uid_index_getpods}
\begin{tabular}{|l|l|}
\hline
\textbf{UID} & \textbf{UID Index} \\
\hline
get-cont-create & 0 \\
\hline
get-cont-getPidsError & 1 \\
\hline
get-cont-getStderrFileInfoError & 2 \\
\hline
get-cont-stderrNotEmpty & 3 \\
\hline
get-cont-completed & 4 \\
\hline
get-cont-running & 5 \\
\hline
\end{tabular}
\end{table}

\subsubsection{The Flowchart for Creating and Monitoring Lifecycle of the Containers in a Pod}

\begin{figure}[htbp]
\centering
\includegraphics[width=\linewidth]{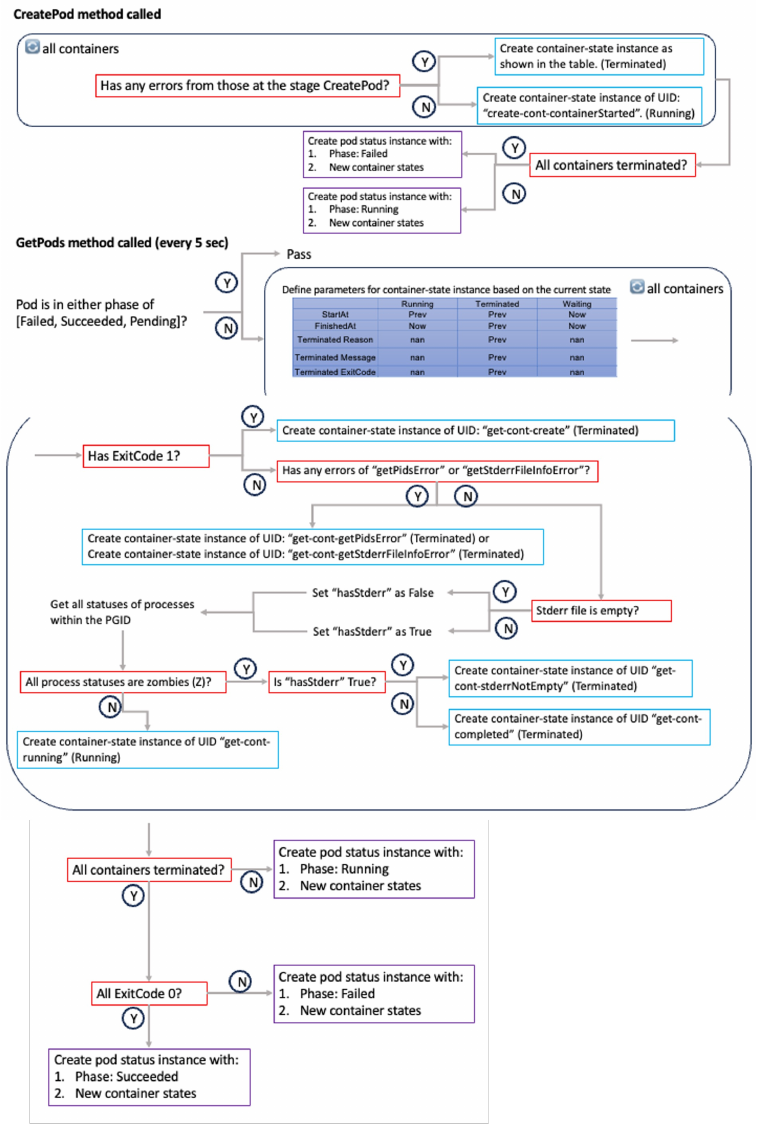}
\caption{Flowchart illustrating the process of creating and monitoring the lifecycle of containers within a pod created by VK. The flowchart includes the initial creation of container states, updating pod status based on container states, and handling errors during the lifecycle. Key blocks indicate looping over containers, creating container states, updating pod status, and redirecting flows based on conditions.}
\label{fig:flowchart}
\end{figure}

The flowchart in Figure \ref{fig:flowchart} illustrates the process of creating and monitoring containers and pods in VK.

\subsubsection{Implementation of Container Lifecycle Management}

To ensure proper lifecycle management, the following Go code initializes and updates pod states based on various transition conditions:

\begin{verbatim}
pod.Status.Conditions = []v1.PodCondition{
  {
    Type:               v1.PodScheduled,
    Status:             v1.ConditionTrue,
    LastTransitionTime: startTime,
  },
  {
    Type:               v1.PodReady,
    Status:             podReady,
    LastTransitionTime: startTime,
  },
  {
    Type:               v1.PodInitialized,
    Status:             v1.ConditionTrue,
    LastTransitionTime: startTime,
  },
}
\end{verbatim}

Similarly, pod readiness is assessed in the \texttt{GetPods} function:

\begin{verbatim}
Conditions: []v1.PodCondition{
  {
    Type:   v1.PodScheduled,
    Status: v1.ConditionTrue,
    LastTransitionTime: *prevPodStartTime,
  },
  {
    Type:   v1.PodInitialized,
    Status: v1.ConditionTrue,
    LastTransitionTime: *prevPodStartTime,
  },
  {
    Type:   v1.PodReady,
    Status: podReady,
    LastTransitionTime: prevContainerStartTime[firstContainerName],
  },
}
\end{verbatim}

\paragraph{Walltime Adjustment}
The system sets the walltime for the container lifecycle as follows:

\begin{verbatim}
sleep $JIRIAF_WALLTIME
echo "Walltime $JIRIAF_WALLTIME has ended. Terminating the processes."
pkill -f "./start.sh"
\end{verbatim}

\subsection{Supporting Horizontal Pod Autoscaling in Kubernetes}
\label{hpa}

For effective implementation of Horizontal Pod Autoscaling (HPA) in Kubernetes, specifically for VK, it is crucial to establish accurate pod conditions. Accurate pod conditions are essential for the optimal functioning of HPA. This section provides essential insights and solutions to achieve this.

\subsubsection{Setting up the Kubernetes Metrics Server}

Prior to implementing the HPA, it is essential to ensure that the metrics server is installed within your Kubernetes cluster. Detailed installation instructions can be found in the official metrics server documentation \cite{metrics-server}.

\subsubsection{Understanding HPA through Code Analysis}

The HPA mechanism critically depends on specific Kubernetes code to evaluate pod readiness, particularly in the context of resource scaling. This essential logic ensures that only ready and appropriately initialized pods are considered for scaling actions based on CPU usage. The following snippet from the Kubernetes source code \cite{k8s-hpa} demonstrates this evaluation process:

\begin{verbatim}
if resource == v1.ResourceCPU {
    var unready bool
    _, condition := podutil.GetPodCondition(&pod.Status, v1.PodReady)
    if condition == nil || pod.Status.StartTime == nil {
        unready = true
    } else {
        if pod.Status.StartTime.Add(cpuInitializationPeriod).After(time.Now()) {
            unready = condition.Status == v1.ConditionFalse || 
                      metric.Timestamp.Before(condition.LastTransitionTime.Time
                      .Add(metric.Window))
        } else {
            unready = condition.Status == v1.ConditionFalse && 
                      pod.Status.StartTime.Add(delayOfInitialReadinessStatus)
                      .After(condition.LastTransitionTime.Time)
        }
    }
    if unready {
        unreadyPods.Insert(pod.Name)
        continue
    }
}
\end{verbatim}

\subsubsection{Implementing Correct Pod Conditions}

For HPA to function as intended, it is crucial to correctly set pod conditions upon creation and accurately update their status based on lifecycle events.

\paragraph{Pod Creation Phase}

The initial conditions for running and failed pods must reflect their true state to avoid misinterpretation by the HPA logic. This setup occurs when calling the method \texttt{CreatePod} (see Section~\ref{lifecycle}). The following details outline the process:

\begin{verbatim}
pod.Status.Conditions = []v1.PodCondition{
  {
    Type:               v1.PodScheduled,
    Status:             v1.ConditionTrue,
    LastTransitionTime: startTime,
  },
  {
    Type:               v1.PodReady,
    Status:             podReady,
    LastTransitionTime: startTime,
  },
  {
    Type:               v1.PodInitialized,
    Status:             v1.ConditionTrue,
    LastTransitionTime: startTime,
  },
}
\end{verbatim}

\paragraph{Pod Retrieving Phase}

\begin{verbatim}
Conditions: []v1.PodCondition{
  {
    Type:   v1.PodScheduled,
    Status: v1.ConditionTrue,
    LastTransitionTime: *prevPodStartTime,
  },
  {
    Type:   v1.PodInitialized,
    Status: v1.ConditionTrue,
    LastTransitionTime: *prevPodStartTime,
  },
  {
    Type:   v1.PodReady,
    Status: podReady,
    LastTransitionTime: prevContainerStartTime[firstContainerName],
  },
}
\end{verbatim}

\subsubsection{HPA Formula Explanation}

The HPA in Kubernetes determines the desired number of pod replicas using the following formula:

\begin{equation}
\texttt{Desired Replicas} = \left\lceil \text{Current Replicas} \times \left( \frac{\text{Current Metric}}{\text{Target Metric}} \right) \right\rceil
\end{equation}

\paragraph{Example}

Assume you have an application deployed with HPA configured to maintain an average CPU utilization of 50\%. If the current average CPU utilization is 90\% and there are 4 current replicas, the desired replicas calculation would be:

\begin{equation}
\texttt{Desired Replicas} = \left\lceil 4 \times \frac{90}{50} \right\rceil = \left\lceil 7.2 \right\rceil = 8
\end{equation}

\subsubsection{Evaluation of HPA using VK}

This section details the procedure for testing the HPA capabilities of pods running on VK. The HPA metric used for this assessment is CPU utilization, as provided by the metrics-server. The source code for this test is available on GitHub \cite{autoscaling-jiriaf}.

\paragraph{Setup}

The test setup includes an HTTP load balancer implemented in Go (\texttt{HPA/load/lb/load\_balancer.go}), which distributes HTTP requests across multiple HTTP servers implemented in Go (\texttt{HPA/load/server.go}). The Kubernetes deployment configuration for these HTTP servers specifies the number of replicas or pods. The HPA manages the scaling of these pods based on the defined metrics.

\paragraph{Load Generation}

HTTP request loads are generated using the \texttt{hey} application \cite{hey}, initiated by the bash script \texttt{HPA/load/add-load.sh}. To install \texttt{hey}, execute the following command:

\begin{verbatim}
go install github.com/rakyll/hey@latest
\end{verbatim}

\paragraph{Test Results}

The test results confirm that the HPA functions effectively with VK, including both the upscaling and downscaling of pods. As the load increases, the HPA scales up the number of pods to manage the additional load. Conversely, when the load decreases, the HPA scales down the number of pods after a five-minute interval from the last scaling operation.

\subsubsection{HTTP Server Deployment and HPA Configuration for Kubernetes}

Here is the deployment file \texttt{HPA/deploy.yaml} for the HTTP server:

\begin{verbatim}
apiVersion: apps/v1
kind: Deployment
metadata:
  name: http-server
spec:
  selector:
    matchLabels:
      app: http-server
  template:
    metadata:
      labels:
        app: http-server
    spec:
      containers:
        - name: http-server
          image: http-server
          command: ["bash"]
          args: [""]
\end{verbatim}

Here is the HPA file \texttt{HPA/hpa.yaml}:

\begin{verbatim}
apiVersion: autoscaling/v2
kind: HorizontalPodAutoscaler
metadata:
  name: http-server-hpa
spec:
  scaleTargetRef:
    apiVersion: apps/v1
    kind: Deployment
    name: http-server
  minReplicas: 1
  maxReplicas: 10
  metrics:
  - type: Resource
    resource:
      name: cpu
      target:
        type: Utilization
        averageUtilization: 30
\end{verbatim}

\subsection{JRM Deployment Using FireWorks}
\label{jrm-deployment}

This section provides an overview of the JRM deployment process, which involves two main components: the FireWorks launchpad and the JRM deployment script. The FireWorks launchpad is a MongoDB database that stores the JRM workflows, while the JRM deployment script is a bash script that sets up and runs a Docker container to launch Slurm jobs for deploying JRMs. For further details and access to the complete code, refer to the GitHub repository \cite{github}.

\subsubsection{Prerequisite: FireWorks Launchpad Setup}

FireWorks is employed to manage the JRM deployment, offering a workflow management system that facilitates the execution of complex workflows. Comprehensive information on FireWorks can be found in \cite{fireworks-url}, and additional guidance is provided by the NERSC introduction to FireWorks \cite{nersc-url}.

\paragraph{Launchpad Configuration File for FireWorks}
The configuration file \texttt{/FireWorks/util/my\_launchpad.yaml} serves as a MongoDB configuration for the FireWorks launchpad. Ensure that the MongoDB port (default is \texttt{27017}) is accessible to the container. If it is not accessible, verify that the port is open to all interfaces.

\paragraph{Setup on the Database Server}
Establish the database and user as specified in the \texttt{my\_launchpad.yaml} file. The \texttt{FireWorks/util/create\_db.sh} script can be utilized to create the database and user.

\paragraph{Setup on the Compute Node}
Prepare the Python environment according to the \texttt{requirements.txt} file. Use the \texttt{FireWorks/util/create\_project.py} script to set up configuration files, generating two files: \texttt{my\_qadapter.yaml} and \texttt{my\_fworker.yaml}. Refer to the example files \texttt{FireWorks/util/my\_launchpad.yaml} and \texttt{FireWorks/util/my\_qadapter.yaml} for guidance. Ensure that MongoDB is accessible from the compute node. If not, consider using SSH tunneling to establish a connection to MongoDB.

\subsubsection{JRM Deployment on Perlmutter at NERSC}

With the FireWorks launchpad set up, the JRM deployment can proceed through the following steps:

\paragraph{Prerequisite: SSH Private Key to NERSC}
A NERSC account and an established private key (e.g., \texttt{\~/.ssh/nersc}) are required for logging into Perlmutter.

\paragraph{Step 1: Create SSH Connections using the Binary \texttt{jrm-create-ssh-connections}}
The binary acts as an HTTP server listening on port \texttt{8888}, creating four SSH connections (db port, apiserver port, jrm port, and custom metrics port) as depicted by the dotted black lines in Figure~\ref{fig:network_map}. Detailed information is available in the \texttt{create-ssh-connections/jrm-fw-create-ssh-connections.go} file. The script looks for available ports for the JRM port (between 10000 and 19999) and custom metrics port (between 20000 and 49999) on the local machine.

\paragraph{Step 2: Configure Environment Variables}
These environment variables are essential for launching JRMs (Job Resource Managers) and establishing SSH tunnels, ensuring seamless integration and communication between different system components.

\begin{verbatim}
nnodes=2
nodetype=cpu
walltime=00:05:00
account=m3792
qos=debug

nodename=vk-nersc-test
site=perlmutter
control_plane_ip=jiriaf2302
apiserver_port=38687
kubeconfig=/global/homes/j/jlabtsai/run-vk/kubeconfig/jiriaf2302
vkubelet_pod_ip=172.17.0.1
jrm_image=docker:jlabtsai/vk-cmd:main
custom_metrics_ports=1234 1423

ssh_remote_proxy=perlmutter
ssh_remote=jlabtsai@128.55.64.13
ssh_key=$HOME/.ssh/nersc
\end{verbatim}

\paragraph{Step 3: Launch JRMs}
Pull the Docker image \texttt{jlabtsai/jrm-fw} from Docker Hub. Follow the above two steps and run the following commands.

\begin{verbatim}
#!/bin/bash
export env_list="env.list"
export jrm_fw_tag="latest"
export logs="$HOME/jrm-launch/logs"

docker run -it --rm --name=jrm-fw -v $logs:/fw/logs --env-file $env_list \
    jlabtsai/jrm-fw:$jrm_fw_tag $@
\end{verbatim}

Within the container, the \texttt{main} directory contains essential files, including:
\texttt{jrm-create-ssh-connections} (a binary), 
\texttt{env.list} (an example setup of environment variables), 
and \texttt{main.sh} (a script for running the image).

\paragraph{Note}
Execute the \texttt{main.sh} script, which accepts the following arguments:
\begin{itemize}
    \item \texttt{add\_wf}: Adds a JRM workflow to the FireWorks database.
    \item \texttt{get\_wf}: Retrieves the table of workflows from the FireWorks database.
    \item \texttt{delete\_wf}: Removes a specific workflow from the FireWorks database.
\end{itemize}

\subsubsection{Port and SSH Tunneling Overview}

\begin{figure}[htbp]
\centering
\includegraphics[width=0.8\textwidth]{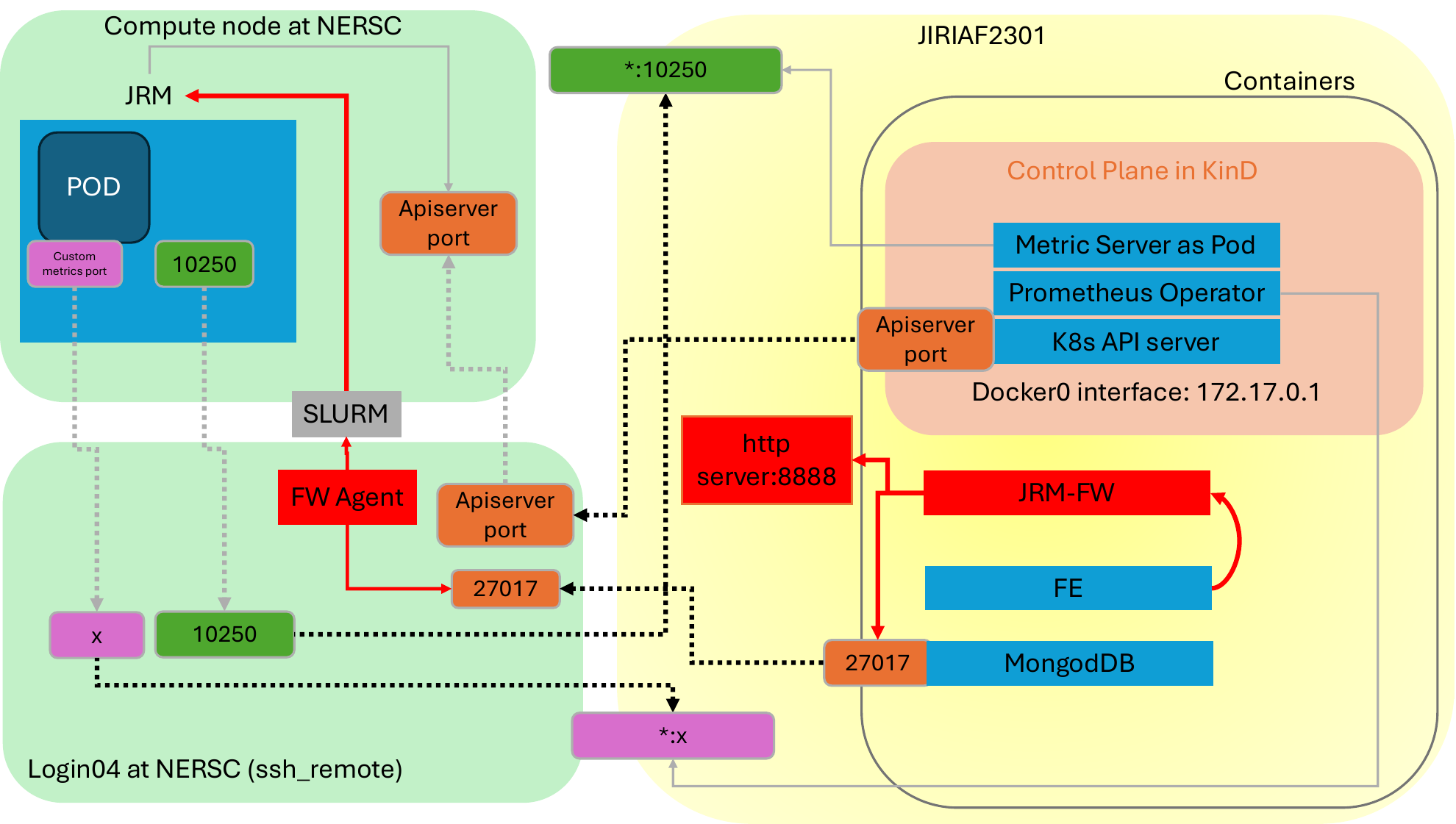}
\caption{
Network map illustrating the ports and SSH tunneling configurations used in the JRM deployment process. Arrowed red lines represent command execution, while black dashed lines indicate SSH tunneling initiated by \texttt{JIRIAF2301}. Solid gray lines represent listening connections, and dashed gray lines represent SSH tunneling from JRM at the compute node to \texttt{login04} at NERSC. The map highlights communication paths for MongoDB, Kubernetes API server, JRM metrics, and custom metrics.
}
\label{fig:network_map}
\end{figure}

Figure~\ref{fig:network_map} illustrates the ports and SSH tunneling used in the JRM deployment process. The Kubernetes cluster in this experiment is deployed using Kubernetes in Docker (KinD) \cite{KinD}.

\subsubsection{Walltime Discrepancy Between JRM and Slurm Job}

The \texttt{JIRIAF\_WALLTIME} variable in \texttt{FireWorks/gen\_wf.py} is intentionally set to be \texttt{60 seconds} less than the walltime of the Slurm job. This adjustment ensures that the JRM has sufficient time to initialize and start running.

Upon the expiration of \texttt{JIRIAF\_WALLTIME}, the JRM will be terminated. The commands for tracking the walltime and terminating the JRM are explicitly defined in the \texttt{FireWorks/gen\_wf.py} file, as illustrated below:

\begin{verbatim}
sleep $JIRIAF_WALLTIME
echo "Walltime $JIRIAF_WALLTIME has ended. Terminating the processes."
pkill -f "./start.sh"
\end{verbatim}

\subsection{Using Prometheus Operator in K8s Cluster to Monitor Application Metrics}
\label{prometheus-operator}

This section explains how to utilize the Prometheus Operator in a Kubernetes (K8s) cluster for monitoring application metrics. The process involves deploying Service and ServiceMonitor objects along with Prometheus instances to collect metrics from the applications.

The Prometheus Operator streamlines the configuration and management of Prometheus monitoring instances within Kubernetes. By leveraging ServiceMonitor objects, users can define how services should be monitored. These configurations enable Prometheus to automatically discover and scrape metrics from specified services.

An important distinction between using a regular kubelet and the virtual-kubelet-cmd lies in the assignment of pod IPs. For pods created by the virtual-kubelet-cmd (VK), the pod IP is defined by the environment variable \texttt{VKUBELET\_POD\_IP} when starting VK. This can result in pods within a Kubernetes deployment sharing the same IP, unlike the unique IPs assigned to pods created by the regular kubelet.

The following subsections provide detailed instructions on setting up the Prometheus Operator, deploying services and ServiceMonitor objects, and configuring Prometheus instances to scrape metrics.

\subsubsection{Setting Up the Prometheus Operator}

The Prometheus Operator can be installed using Helm or by deploying the necessary components from the kube-prometheus project. The kube-prometheus project provides a comprehensive set of manifests for deploying and configuring Prometheus, Alertmanager, and Grafana, along with the necessary configuration for scraping metrics from Kubernetes services \cite{kube-prometheus}.

\subsubsection{Deploying Applications and Monitoring with Unique Pod IPs}

Due to the different pod IPs of pods created by VK, we describe how to deploy an application and its monitoring depending on the uniqueness of the pod IP.

\begin{figure}[htbp]
\centering
\includegraphics[width=\textwidth]{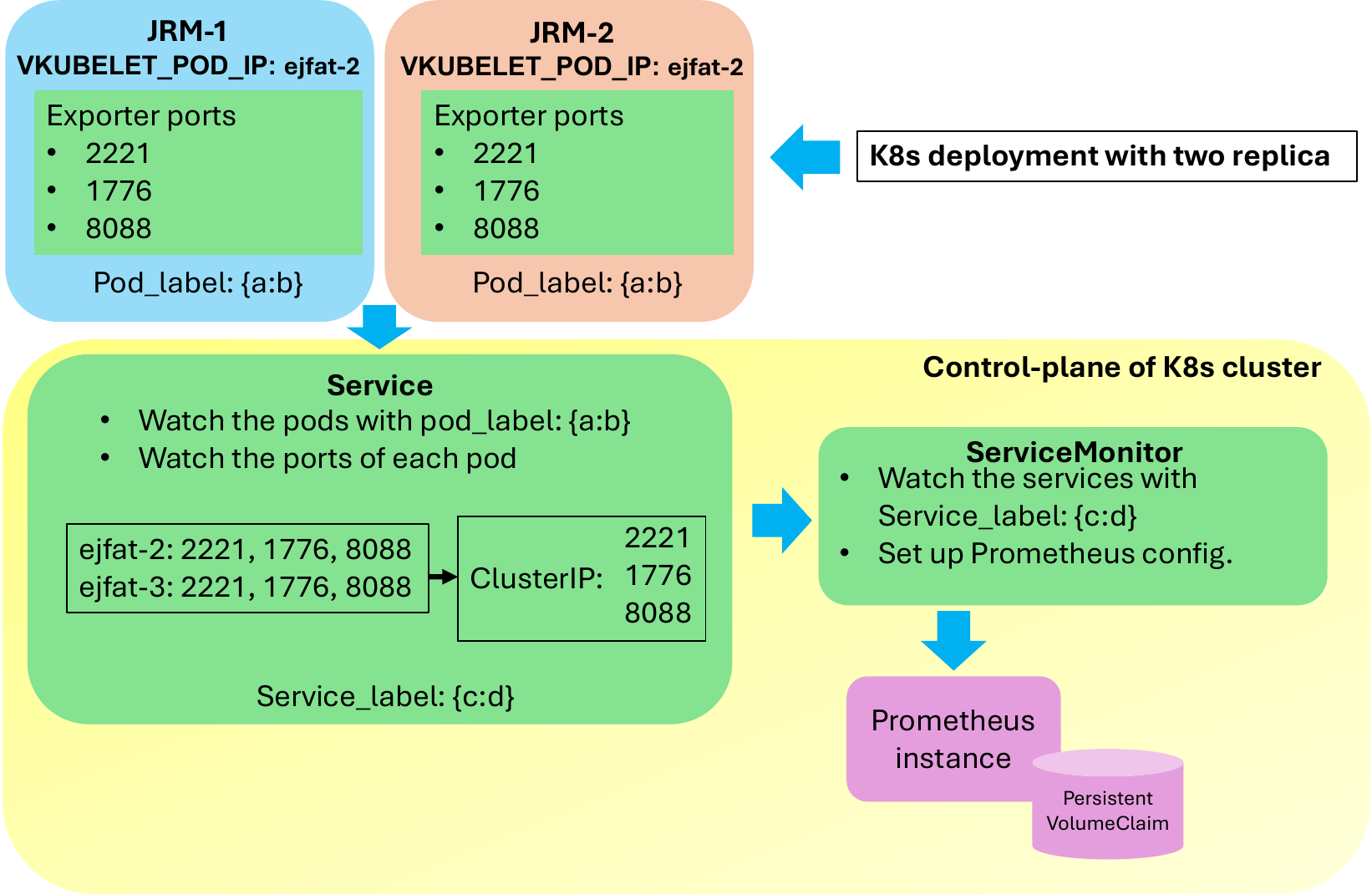}
\caption{Deployment and monitoring of Kubernetes applications with unique pod IPs using Prometheus Operator. The figure illustrates a Kubernetes application with two replicas, each assigned a unique IP address based on the DNS name of the node it runs on. Metrics are exported from each pod on ports 2221, 1776, and 8088, aggregated by a Service object, and monitored by a ServiceMonitor for Prometheus scraping.}
\label{fig:unique-pod-ips}
\end{figure}

\paragraph{Pod IP and Metrics Ports}
Figure \ref{fig:unique-pod-ips} illustrates the deployment of a Kubernetes application with two replicas, resulting in the creation of two pods. These pods are scheduled to run on nodes with the DNS names `ejfat-2` and `ejfat-3`, which are set up on the control-plane. Each pod's IP address is defined by the `VKUBELET\_POD\_IP` environment variable (see Table \ref{table:environment_variables}) when starting the virtual kubelet (VK). This ensures that each pod is assigned a specific IP address corresponding to the DNS name of the node it is running on. For example, the pod running on the node with DNS name `ejfat-2` will have the IP address `ejfat-2`, and the pod on the node with DNS name `ejfat-3` will have the IP address `ejfat-3`. Each pod exports metrics on the same set of ports: 2221, 1776, and 8088. The pods have labels `a:b` for identification.

\paragraph{Service Creation}
A Kubernetes Service object is created to expose the metrics endpoints of these pods. The Service object aggregates the exported ports from both pods, creating a unified service that maps `ejfat-2:1776` and `ejfat-3:1776` along with the other ports. Since the service is set up on the control-plane (CP) of the cluster, ensure that `ejfat-2:ports` and `ejfat-3:ports` (where ports are 2221, 1776, and 8088) can be reached from the CP. This Service is responsible for combining these endpoints under a single ClusterIP, which allows Prometheus to access the metrics uniformly. The Service watches pods with the label `a:b`. Ensure that the labels of monitored pods are set up correctly here.

\paragraph{ServiceMonitor Configuration}
A ServiceMonitor object is defined to monitor the Service. The ServiceMonitor looks at the specified Services and configures Prometheus to scrape the metrics from these endpoints. The ServiceMonitor watches services with the label `c:d`. Ensure that the labels of the monitored services are set up correctly to allow Prometheus to discover and scrape the metrics properly.

\paragraph{Prometheus Instance}
The Prometheus instance, configured through the ServiceMonitor, scrapes the metrics from the Service endpoints. The data is then stored in a PersistentVolume Claim for durable storage and further analysis.

\subsubsection{Deploying Applications and Monitoring with Same Pod IPs}

When the pod IPs are the same, it creates a different challenge. If the exporter ports are also the same for each pod, the service will not be able to route the ports correctly due to the identical pod IPs.

\paragraph{Pod IP and Metrics Ports}
In this scenario, both pods are assigned the same IP address by the `VKUBELET\_POD\_IP` environment variable. Since the IPs and ports are identical, this can cause conflicts in routing the metrics.

\paragraph{Service Creation}
To resolve this issue, map the exporter ports from the pods onto different ports for each pod on the control-plane (CP). Create separate services for each pod, ensuring that each service maps the pod's exporter ports to unique ports on the CP. This allows the metrics to be correctly routed despite the identical pod IPs.

\paragraph{ServiceMonitor Configuration}
Create separate ServiceMonitor objects for each service. Each ServiceMonitor will watch its corresponding service and set up Prometheus to scrape metrics from the unique ports. Ensure that the labels for these services are correctly configured so that Prometheus can discover and scrape the metrics properly.

\paragraph{Prometheus Instance}
Use the same Prometheus instance to scrape metrics from both services. This configuration allows the metrics to be collected together and analyzed as a unified dataset.

\section{Proof of Concept}
\begin{figure}[h]
    \centering
    \includegraphics[width=\textwidth]{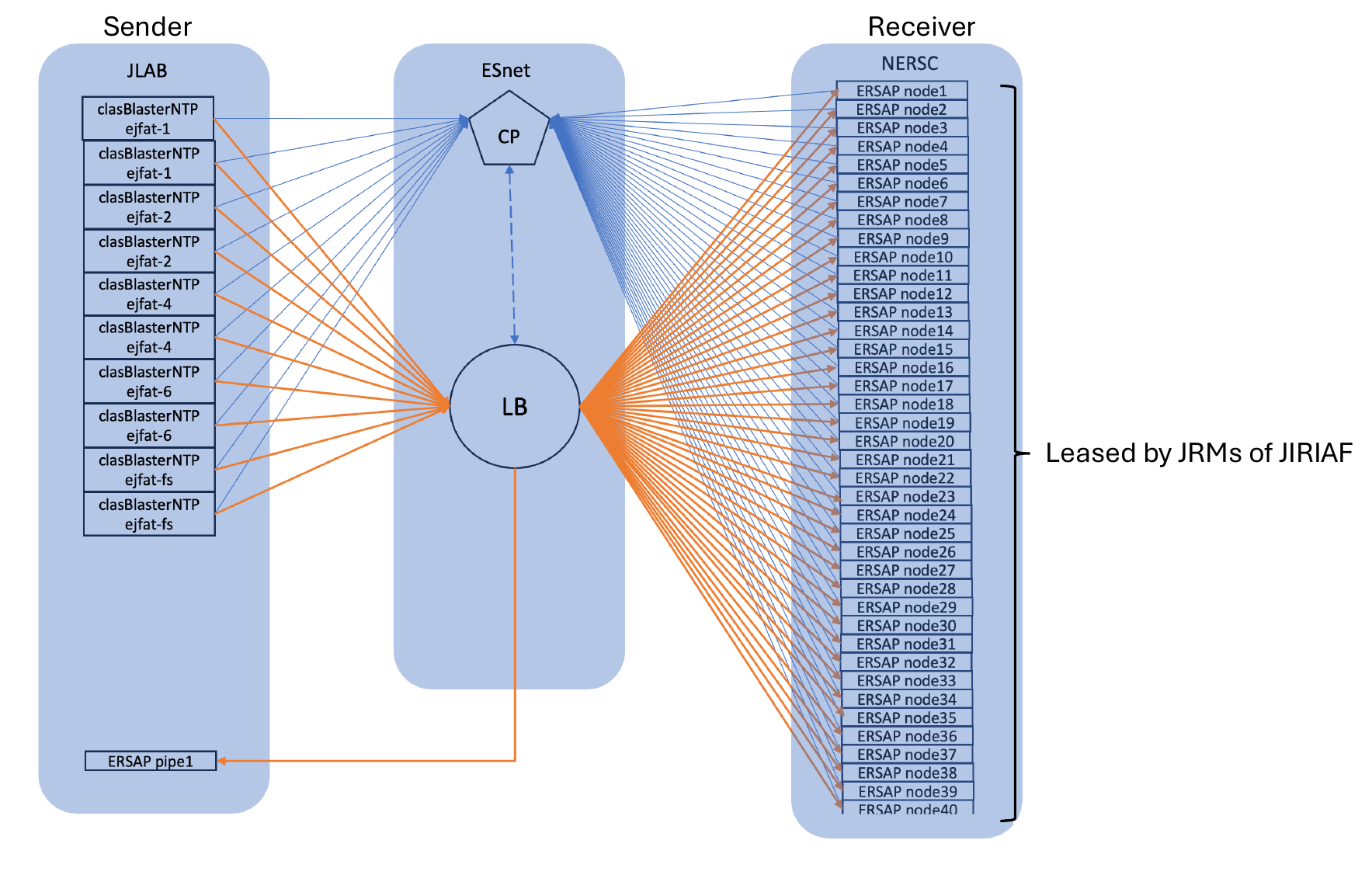}
    \caption{The ERSAP framework utilized in the JIRIAF deployment on the Perlmutter compute nodes at NERSC. The CLAS12 event reconstruction application ran on each node in the JIRIAF Kubernetes cluster, demonstrating the JIRIAF deployment's effectiveness in handling high-volume data-stream processing.}
    \label{fig:ersap_framework}
\end{figure}

A 40-node reservation on the Perlmutter system at NERSC was activated to deploy data-stream processing pipelines. This deployment utilized the JIRIAF framework across the JIRIAF Kubernetes cluster nodes, each executing the CLAS12 event reconstruction application. This application was optimized to fully leverage all available processing cores within the ERSAP framework (see Figure~\ref{fig:ersap_framework}).

To demonstrate the effectiveness of JIRIAF, a proof of concept was conducted using the CLAS12 experiment. Event streams were transmitted to the NERSC computing facility via the EJFAT transport system. JRMs/VKs were deployed on 40 nodes within the NERSC cluster for stream processing workflows. The ERSAP workflow was deployed for CLAS12 reconstruction.

JRMs/VKs of JIRIAF as agents were deployed by the SLURM batch job system at NERSC. These JRMs formed K8s nodes waiting for deployments. The ERSAP processing application was containerized and uploaded to the Shifter container hub at NERSC. A K8s deployment YAML file was created and applied to the K8s API server on the control-plane at JLAB (refer to \hyperref[deploying-pods]{Deploying and Managing Pods on Virtual-Kubelet-Cmd Nodes}). The monitoring system scraped and stored metrics data on the control-plane at JLAB (refer to \hyperref[prometheus-operator]{Using Prometheus Operator in K8s Cluster to Monitor Application Metrics}). Notice that we deployed an independent Prometheus server outside our Kubernetes cluster during this test.

Concurrent with the deployment, CLAS12 raw event data began streaming from JLAB, generating traffic volumes exceeding 100 Gbps directed towards the ESnet Load Balancer (LB). This marked the commencement of the operational functionality of the distributed remote data stream processing.

\begin{figure}[h]
    \centering
    \includegraphics[width=\textwidth]{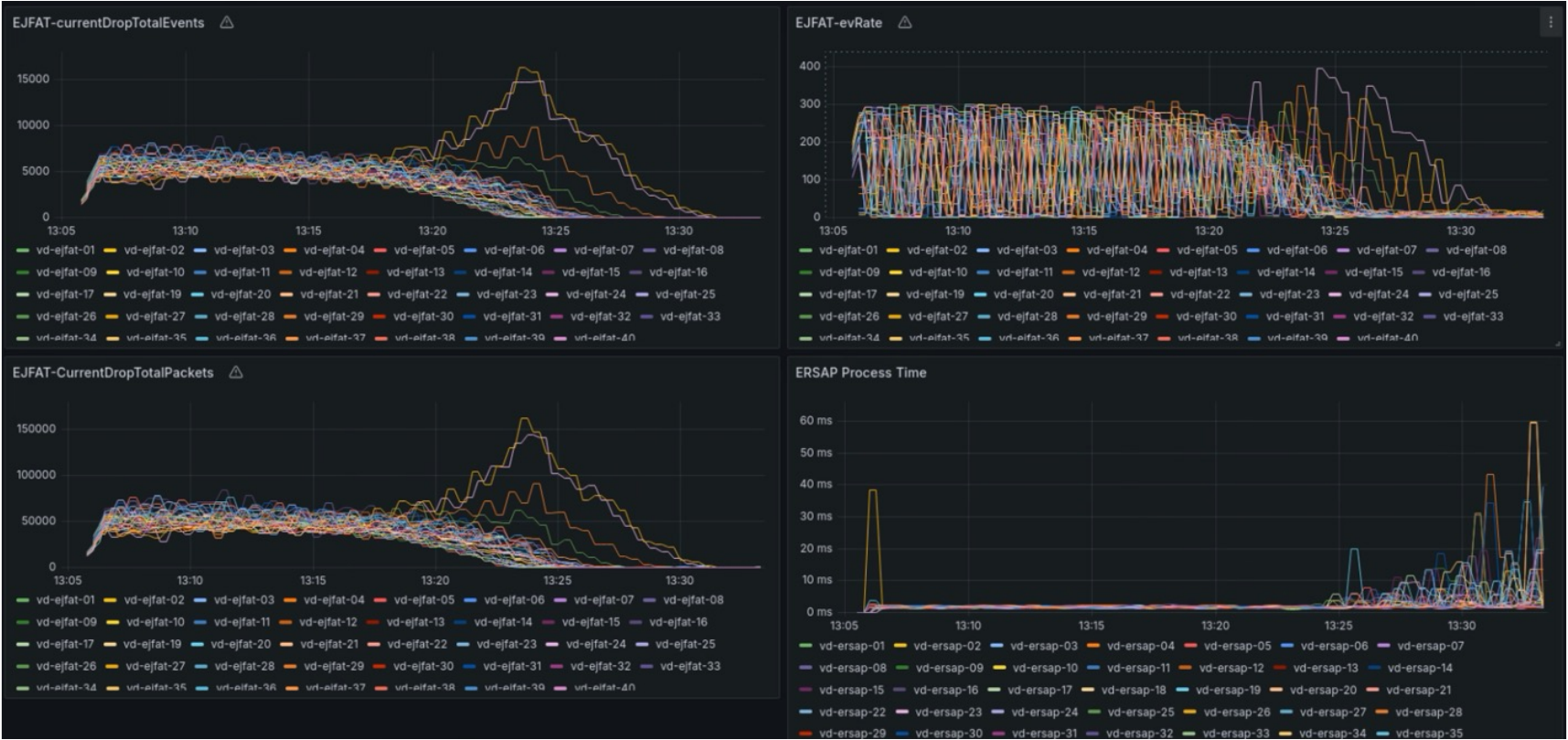}
    \caption{Monitoring system metrics scraped from applications during the JIRIAF deployment. The figure shows the metrics collected by the monitoring system, providing insights into the performance and resource utilization of the deployed CLAS12 event reconstruction application across the NERSC cluster nodes. These metrics are crucial for evaluating the effectiveness and efficiency of the JIRIAF framework in a high-performance computing environment.}
    \label{fig:monitoring_metrics}
\end{figure}

The monitoring system metrics scraped from applications during the JIRIAF deployment are shown in Figure~\ref{fig:monitoring_metrics}.

\newpage

\subsection{Deploying JRM/VK Nodes via SLURM}
A 40-node reservation on the Perlmutter system was activated for deploying data-stream processing pipelines using the JIRIAF framework.

The deployment of these JRMs/VKs was automated through a SLURM script named \texttt{nersc-slurm.sh}. This script handles the deployment across multiple nodes efficiently by leveraging SLURM's job scheduling capabilities.

\begin{verbatim}
#!/bin/bash

# SBATCH -N 40
# SBATCH -C cpu
# SBATCH -q regular
# SBATCH -J 100g
# SBATCH -t 03:00:00
# SBATCH --reservation=100g

for i in $(seq 1 40)
do
    i_padded=$(printf "%02d" $i)
    echo $i_padded
    srun -N1 /global/homes/j/jlabtsai/run-vk/slurm/mylin/node-setup.sh $i_padded &
    sleep 3
done
wait
\end{verbatim}

This SLURM script configures and submits a job to reserve 40 CPU-based nodes. It then executes the \texttt{node-setup.sh} script on each node in a staggered manner to prevent overloading the system's resources.

Additionally, each Virtual Kubelet node is configured using the \texttt{node-setup.sh} script, which sets up environment variables, configures SSH tunnels for secure communication, and prepares the node for the deployment of monitoring tools and applications.

\begin{verbatim}
#!/bin/bash

export CONTROL_PLANE_IP="jiriaf2302"
export APISERVER_PORT="38687"
export NODENAME="vk-nersc$1"
export KUBECONFIG="/global/homes/j/jlabtsai/run-vk/kubeconfig/$CONTROL_PLANE_IP"
export VKUBELET_POD_IP="172.17.0.1"
export KUBELET_PORT="100"$1

export JIRIAF_WALLTIME="0"
export JIRIAF_NODETYPE="cpu"
export JIRIAF_SITE="nersc"

export proxy_remote="jlabtsai@128.55.64.25"

ssh -NfL $APISERVER_PORT:localhost:$APISERVER_PORT $proxy_remote
ssh -NfR $KUBELET_PORT:localhost:$KUBELET_PORT $proxy_remote

export ersap_exporter="200"$1
export process_exporter="300"$1
export ejfat_exporter="400"$1

ssh -NfR $ersap_exporter:localhost:2221 $proxy_remote
ssh -NfR $process_exporter:localhost:1776 $proxy_remote
ssh -NfR $ejfat_exporter:localhost:8080 $proxy_remote

shifter --image=docker:jlabtsai/vk-cmd:main -- /bin/bash -c "cp -r /vk-cmd `pwd`/$NODENAME"
cd `pwd`/$NODENAME
./start.sh $KUBECONFIG $NODENAME $VKUBELET_POD_IP $KUBELET_PORT $JIRIAF_WALLTIME \
    $JIRIAF_NODETYPE $JIRIAF_SITE
\end{verbatim}

These scripts together facilitate a robust and scalable deployment strategy for JRMs/VKs, crucial for managing high-performance computing tasks across distributed environments.

\subsection{Deploying ERSAP Applications}
The ERSAP applications are deployed on the 40 JRM/VK nodes using Kubernetes Helm charts \cite{helm} to manage the deployment of containers. The deployment utilizes a series of ConfigMaps and Pods defined in Helm templates.

\paragraph{Helm Deployment Commands}
The following script deploys the ERSAP applications on each JRM/VK node using Helm:

\begin{verbatim}
for i in $(seq 1 40)
do
    i_padded=$(printf "%02d" $i)
    echo $i_padded
    helm install ersap$i ersap --set name=$i_padded
done
\end{verbatim}

This script sequentially deploys the Helm chart on each of the 40 nodes, adjusting the configuration dynamically.

\section{Digital Twin for Simulated Stream Processing System}
In the broader context of our study on optimizing computational resource allocation in high-throughput systems, we integrated a digital twin model to enhance real-time monitoring and control capabilities. The digital twin component leverages a Dynamic Bayesian Network (DBN) to simulate the behavior of a queue system, providing valuable insights into system dynamics and aiding in decision-making processes. We utilized the code from \cite{kapteyn2021probabilistic} to build our DBN model, demonstrating the practical application of their proposed framework.

\subsection{Digital Twin Model and Methodology}

The digital twin model was developed to mirror the state and behavior of a physical queue system, comprising a stream sender and receiver with a FIFO queue. The DBN framework was employed to capture dependencies among system variables, offering a probabilistic approach to real-time data assimilation and state estimation. 

Our experimental setup involved adjusting the event sending rates (\(\lambda\)) and measuring the resulting processing rates (\(\mu\)) and observed queue lengths (Obs. \(L_q\)) under different computational capacities (16 and 32 threads). The theoretical queue length (Calc. \(L_q\)) was calculated using the M/M/1 queue theory equation:

\begin{align}
L_q = \frac{\lambda^2}{\mu (\mu - \lambda)}
\label{eq:Lq_sec}
\end{align}

The data collected from these experiments were used to construct and validate the DBN model, enabling it to make accurate state predictions and recommend optimal control actions.

\begin{figure}[htbp]
\centering
\includegraphics[width=3.5in]{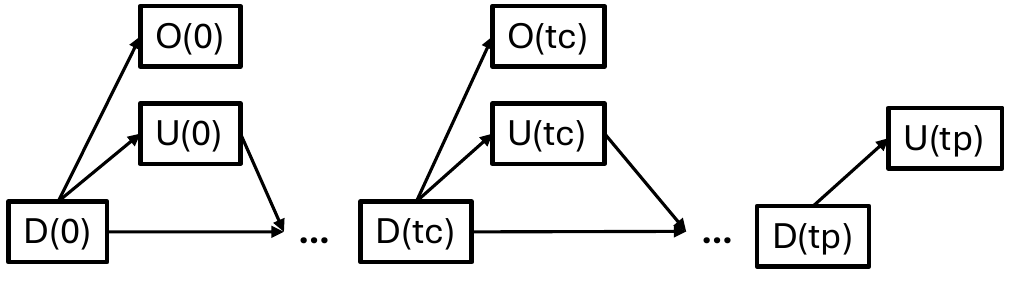}
\caption{Dynamic Bayesian Network representation of the digital twin. It consists of the nodes \(D(t)\), \(O(t)\), and \(U(t)\).}
\label{fig:bayesian_network_sec}
\end{figure}

The DBN structure is depicted in Figure~\ref{fig:bayesian_network_sec}, illustrating the relationships between the digital twin state (\(D(t)\)), control (\(U(t)\)), and observation (\(O(t)\)) nodes.

\subsection{Ground Truth and Experimental Data}

To evaluate the digital twin's performance, we defined the ground truth behavior of the queue system over time using a piecewise function. This artificial ground truth allowed us to control the state changes in a predictable manner:

\begin{itemize}
    \item If timestep is less than 10, state increases by 0.4.
    \item If timestep is between 20 and 30, state decreases by 0.4.
    \item If timestep is between 40 and 50, state increases by 0.4.
    \item If timestep is between 60 and 70, state decreases by 0.4.
    \item No changes in state during unspecified periods.
\end{itemize}

The observation data \(o_t\) as Obs. \(L_q\) at timestep \(t\) were constructed by interpolating data from Tables~\ref{tab:16_threads_sec} and \ref{tab:32_threads_sec}, which highlight the experimental setup.

\begin{table}[htbp]
\caption{System Metrics for 16 Threads}
\centering
\small
\begin{tabular}{|c|c|c|c|c|c|}
\hline
State & \(\lambda\) (Hz) & \(\mu\) (Hz) & Proc. Units & Obs. \(L_q\) & Calc. \(L_q\) \\
\hline
0 & 162 & 167 & 16 & 32 & 33.74 \\
1 & 163 & 167 & 16 & 41 & 43.48 \\
2 & 164 & 167 & 16 & 58 & 60.52 \\
3 & 165 & 167 & 16 & 97 & 98.01 \\
4 & 166 & 167 & 16 & 241 & 248.00 \\
\hline
\end{tabular}
\label{tab:16_threads_sec}
\end{table}

\begin{table}[htbp]
\caption{System Metrics for 32 Threads}
\centering
\small
\begin{tabular}{|c|c|c|c|c|c|}
\hline
State & \(\lambda\) (Hz) & \(\mu\) (Hz) & Proc. Units & Obs. \(L_q\) & Calc. \(L_q\) \\
\hline
0 & 162 & 222 & 32 & 1.56 & 1.96 \\
1 & 163 & 222 & 32 & 2.5 & 2.02 \\
2 & 164 & 222 & 32 & 2.56 & 2.08 \\
3 & 165 & 222 & 32 & 3.5 & 2.14 \\
4 & 166 & 222 & 32 & 3.56 & 2.21 \\
\hline
\end{tabular}
\label{tab:32_threads_sec}
\end{table}

\subsection{Control History and Observations}

\begin{figure}[htbp]
\centering
\includegraphics[width=3.5in]{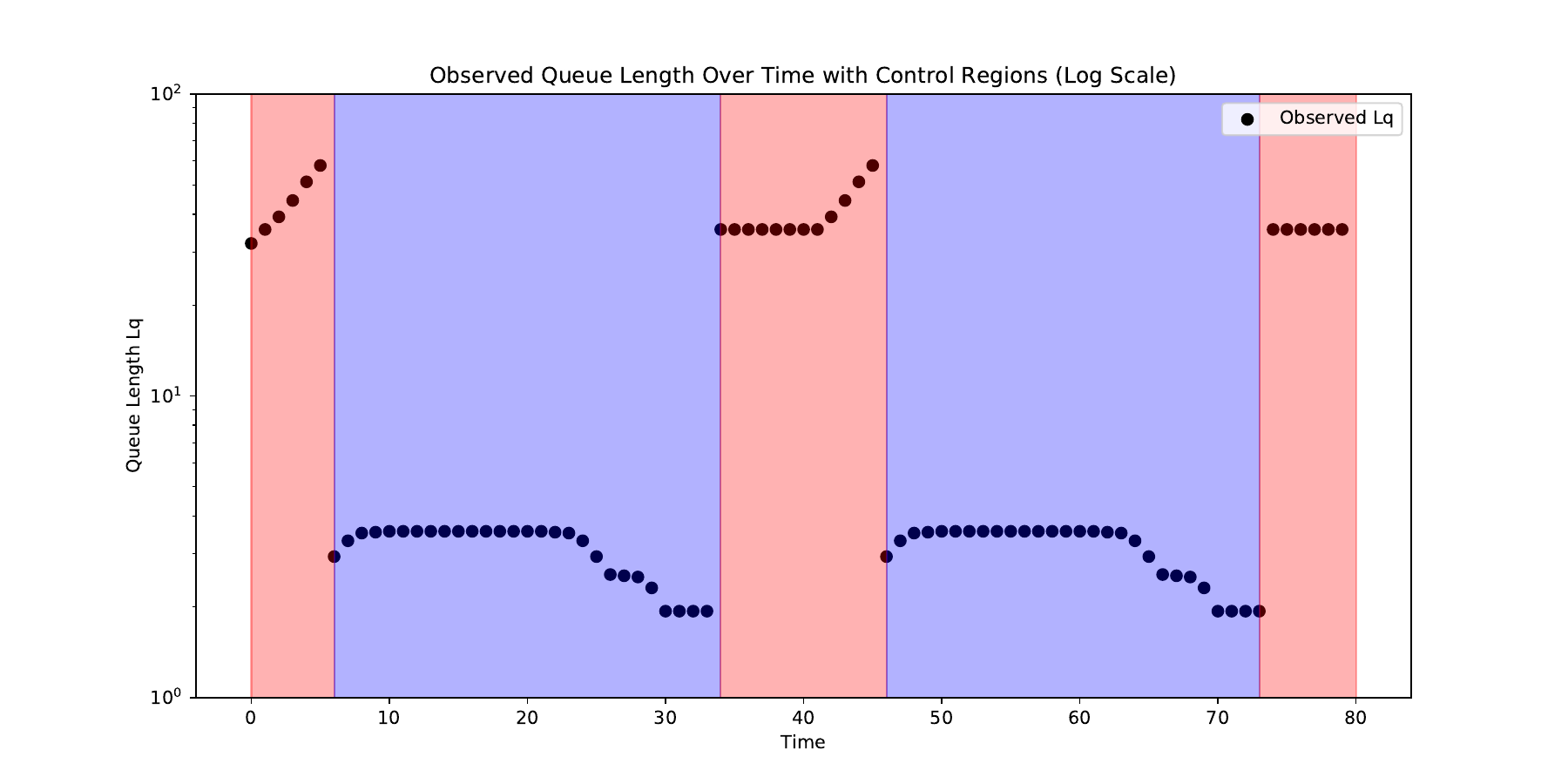}
\caption{Observed Queue Length Over Time with Control Regions (Log Scale). Black dots represent observed queue length, while blue and red regions indicate different control actions.}
\label{fig:queue_length_sec}
\end{figure}

As shown in Figure~\ref{fig:queue_length_sec}, the queue system initially operates with 16 threads, which results in longer queue lengths. As pressure increases, the digital twin recommends switching to 32 threads to reduce system congestion.

\begin{figure}[htbp]
\centering
\includegraphics[width=3.5in]{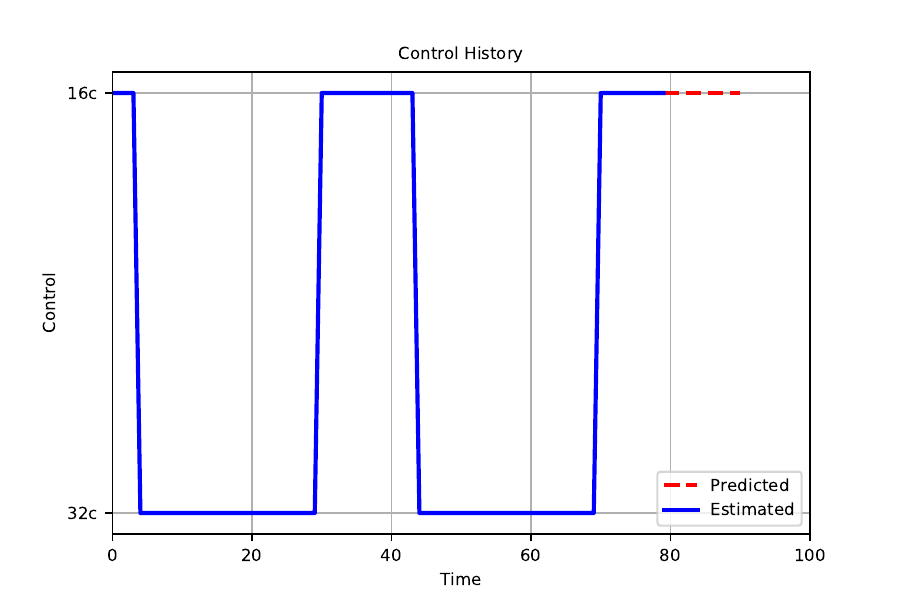}
\caption{Control History Over Time. The red dashed line shows predicted control actions, while the blue line represents estimated control actions.}
\label{fig:control_history_sec}
\end{figure}

As demonstrated in Figure~\ref{fig:control_history_sec}, the estimated control actions closely followed the predicted control recommendations, demonstrating the model's capability to adaptively manage system resources.

\subsection{Challenges and Future Directions}

While the digital twin model demonstrated robust performance, certain challenges were identified. The model exhibited delays in state tracking during periods of decreasing queue lengths, which may be attributed to indistinguishable calculated queue lengths and inaccuracies in the Conditional Probability Tables (CPTs). Addressing these challenges is crucial for enhancing predictive accuracy.

Future work will focus on optimizing the CPTs using historical data and refining the DBN model to handle more complex scenarios. Additionally, we plan to integrate the digital twin with other resource allocation frameworks to provide a holistic optimization solution.

\subsection{Conclusion}

The digital twin model, as part of our broader study, has proven to be a powerful tool for real-time monitoring and control of queue systems. By integrating the DBN framework, we have achieved significant improvements in system performance and efficiency. The insights gained from this integration will inform future developments in digital twin technology for high-throughput computational environments.

\section*{Acknowledgements}
This project is funded through the Thomas Jefferson National Accelerator Facility LDRD program. This material is based upon work supported by the U.S. Department of Energy Office of Science Office of Nuclear Physics under contract DE-AC05-06OR23177.

\bibliographystyle{plain}
\bibliography{references}

\end{document}